\documentclass{aa}
\usepackage{graphicx} 
\usepackage{times,float}

\begin{document}

\title{\object{GRB  050509b}:  the elusive  optical/nIR/mm  afterglow of  a
short-duration  GRB\thanks{Based  on  observations  taken with  the  1.2  m
Mercator telescope at Observatorio del Roque de los Muchachos, with the 1.5
m telescope at  Observatorio de Sierra Nevada, with the 2.2 m and 3.5 m 
telescopes at
the Centro  Astron\'omico Hispano Alem\'an  (CAHA) at Calar  Alto (operated
jointly by  the Max-Planck Institut  f\"ur Astronomie and the  Instituto de
Astrof\'{\i}sica de Andaluc\'{\i}a (CSIC)) and  with the 6.0 m telescope at
the Special Astrophysical Observatory in Russia.}  }

\author{A. J.    Castro-Tirado      \inst{1} 
   \and A.       de Ugarte Postigo  \inst{1}
   \and J.       Gorosabel          \inst{1}
   \and T.       Fathkullin         \inst{2} 
   \and V.       Sokolov            \inst{2}
   \and M.       Bremer             \inst{3}
   \and I.       M\'arquez          \inst{1}
   \and A.~J.~   Mar\'{\i}n         \inst{1}
   \and S.~      Guziy              \inst{1,4}
   \and M.       Jel\'{\i}nek       \inst{1}
   \and P.       Kub\'anek          \inst{5}
   \and R.       Hudec              \inst{5}
   \and S.       Vitek              \inst{6}
   \and T. J.    Mateo Sanguino     \inst{7}
   \and A.       Eigenbrod          \inst{8} 
   \and M.~D.~   P\'erez-Ram\'{\i}rez \inst{9}
   \and A.~      Sota               \inst{1}
   \and J.       Masegosa           \inst{1}
   \and F.       Prada              \inst{1} 
   \and M.       Moles              \inst{1} 
      }

\offprints{A.J. Castro-Tirado, \email{ajct@iaa.es} } 
 
 \institute{Instituto de Astrof\'\i sica de Andaluc\'\i a (IAA-CSIC), P.O. Box 3.004, E-18.080 Granada, Spain. 
           %username@iaa.es 
       \and Special Astrophysical Observatory, Russian Academy of Sciences, Karanchai-Cherkessia, Nizniy-Arkhyz, 357147, Russia. 
           %username@sao.ru 
       \and Institute de Radioastronomie Milimetrique (IRAM), 300 rue de la 
            Piscine, 38406 Saint Martin d'H\'eres, France. 
           %username@iram.fr 
       \and Nikolaev State University, Nikolskaya 24, 54.030 Nikolaev, Ukraine. 
           %username@iaa.es 
       \and Astronomical Institute, Academy of Sciences of the Czech Republic, 
            25165 Ond\v rejov, Czech Republic.
           %username@iaa.es
       \and Fakulta electrotechnick\'a, Czech Technical University, 121 35 
            Praha, Czech Republic.
           %username@iaa.es
       \and Dept. de Ing. Electr\'onica, Sistemas Inform\'aticos y 
            Autom\'atica, Univ. de Huelva, 21.819 Palos de la Frontera (Huelva),
            Spain.
           %username@iaa.es
       \and Ecole Polytechnique F\'ed\'erale de Lausanne, Lab. 
            d'Astrophysique, Observatoire,  CH-1290 Chavannes-des-Bois, 
            Switzerland. 
           %username@efl.ch
       \and  Departamento de F\'{\i}sica (EPS), Univ. de Ja\'en, Campus Las Lagunillas, E-23,071 Ja\'en, Spain.
           }

\date{Received / Accepted } 
 
\abstract{We present  multiwavelength (optical/near infrared/millimetre) 
  observations  of a
  short  duration  gamma-ray burst  detected  by  {\it Swift}  (\object{GRB
  050509b}) collected between 0 seconds  and $\sim$  18.8 
  days  after the event.  No
  optical, near infrared or millimetre  emission has been detected in spite
  of the well localised  X-ray  afterglow, confirming
  the  elusiveness of  the  short  duration events.   We  also discuss  the
  possibility of  the burst  being located in  
  %the cluster of  galaxies NSC J123610+285901  
  a cluster of galaxies at  $z$ =  0.225  or beyond.   
  In  the  former case,  the
  spectral energy distribution of the neighbouring, potential host galaxy,
  favours a system harbouring an  evolved dominant stellar
  population  (age $\sim360$  Myr), unlike most long duration GRB host 
  galaxies observed so far, i.e. thus giving support to a compact binary 
  merger origin. Any underlying supernova that could be associated 
  with this particular event should have been at least 3 magnitudes fainter 
  than the type Ib/c SN 1998bw and 2.3 magnitudes fainter 
  than a typical type Ia SN.   
  }
 
\maketitle 
 
\section{Introduction} 
 
Gamma Ray Bursts  (GRBs) are flashes of high  energy ($\sim$1~keV--10 GeV)
photons (Fishman and Meegan 1995), occurring at cosmological distances. The
detection  of counterparts  at other  wavelengths for  long  duration GRBs,
revealed  their cosmological origin  (see van  Paradijs et  al. 2000  for a
review) and it  is accepted nowadays that they are  associated with 
the death of massive stars.  
On the other hand, the nature of  short duration GRBs, a
class  that  comprises  about 25\%  of  all  events  (Mazets et  al.  1981;
Kouveliotou et  al. 1993),  still remains a  puzzle. No  counterparts have
been discovered so far, in spite  of intense efforts in order to detect the
optical,  infrared and  radio counterparts  to several  short,  hard bursts
(Kehoe et  al. 2001; Gorosabel  et al. 2002;  Hurley et al. 2002;  Klotz et
al. 2003). A possible optical transient, related to GRB 000313 was proposed
by  Castro-Tirado  et   al.  (2002)  but  a  firm   conclusion  could  
not be established.

\begin{table*} 
      \begin{center} 
            \caption{Journal of optical and nIR observations of the \object{GRB 050509b} field.} 
                     \begin{tabular}{@{}lccccc@{}} 
 
{\scriptsize Date of 2005 UT (mid exposure)} & {\scriptsize Telescope/Instrument} & {\scriptsize Filtre} & {\scriptsize Exposure Time (s)} & {\scriptsize Seeing (arcsec)}   & {\scriptsize Limiting Magnitude (3$\sigma$)} \\ 
%{\scriptsize (mid exposure)}  & {\scriptsize Instrument} &        &    {\scriptsize (seconds)}  & {\scriptsize (arcsec)} & {\scriptsize Magnitude} \\  
%{\scriptsize Date of 2005 UT} & {\scriptsize Telescope/} & {\scriptsize Filtre} & {\scriptsize Exposure Time} & {\scriptsize Seeing}   & {\scriptsize Limiting} \\ 
%{\scriptsize (mid exposure)}  & {\scriptsize Instrument} &        &    {\scriptsize (seconds)}  & {\scriptsize (arcsec)} & {\scriptsize Magnitude} \\  
\hline 
{\scriptsize May 09,  04:00.25} & {\scriptsize BOOTES-1 (ASCCD)} & {\scriptsize $R$} &       {\scriptsize 30}   & {\scriptsize 3.0}    &   {\scriptsize 6.0}   \\
{\scriptsize May 09,  04:40}   & {\scriptsize 1.2Mer (MEROPE)}  & {\scriptsize $R$} &      {\scriptsize 600}   & {\scriptsize 2.5}    & {\scriptsize 21.0}   \\ 
{\scriptsize May 09,  23:40}   & {\scriptsize 1.2Mer (MEROPE)}  & {\scriptsize $R$} &      {\scriptsize 600}   & {\scriptsize 1.3}    &  {\scriptsize 22.1}   \\
{\scriptsize May 11, 20:00}   & {\scriptsize 6.0BTA (SCORPIO)} & {\scriptsize $R$} &   {\scriptsize 7\,020}   & {\scriptsize 2.0}    &  {\scriptsize 26.0}   \\
{\scriptsize May 18, 23:00}   & {\scriptsize 1.5OSN (CCD)}     & {\scriptsize $I$} &   {\scriptsize 3\,600}   & {\scriptsize 1.4}    &  {\scriptsize 21.9}   \\
{\scriptsize May 27,  22:57} & {\scriptsize 2.2CAHA (BUSCA)}        & {\scriptsize $R$} &   {\scriptsize 2\,500}   & {\scriptsize 1.4}    &  {\scriptsize 23.9}   \\ 
\hline
{\scriptsize May 09,   21:21}  & {\scriptsize 3.5CAHA (OMEGA2000)}    & {\scriptsize $H$} &   {\scriptsize 3\,378}   & {\scriptsize 2.1}    &  {\scriptsize 20.2}   \\
{\scriptsize May 09,   23:22}  & {\scriptsize 3.5CAHA (OMEGA2000)}    & {\scriptsize $K$} &   {\scriptsize 2\,730}   & {\scriptsize 1.6}    &  {\scriptsize 18.8}   \\ 
{\scriptsize May 10,  00:05}  & {\scriptsize 3.5CAHA (OMEGA2000)}    & {\scriptsize $J$} &   {\scriptsize 3\,656}   & {\scriptsize 2.1}    &  {\scriptsize 21.2}   \\ 
{\scriptsize May 10,  02:29}  & {\scriptsize 3.5CAHA (OMEGA2000)}    & {\scriptsize $H$} &   {\scriptsize 3\,563}   & {\scriptsize 2.4}    &  {\scriptsize 20.5}   \\ 
{\scriptsize May 10,  21:50}  & {\scriptsize 3.5CAHA (OMEGA2000)}    & {\scriptsize $J$} &   {\scriptsize 1\,897}   & {\scriptsize 1.9}    &  {\scriptsize 20.3}   \\ 
{\scriptsize May 12,  21:55}  & {\scriptsize 3.5CAHA (OMEGA2000)}    & {\scriptsize $H$} &   {\scriptsize 2\,314}   & {\scriptsize 1.7}    &  {\scriptsize 20.3}   \\
{\scriptsize May 15,  22:29}  & {\scriptsize 3.5CAHA (OMEGA2000)}    & {\scriptsize $J$} &   {\scriptsize 1\,620}   & {\scriptsize 1.4}    &  {\scriptsize 21.7}   \\
{\scriptsize May 17,  23:00}  & {\scriptsize 3.5CAHA (OMEGA2000)}    & {\scriptsize$J$} &   {\scriptsize3\,656}   & {\scriptsize2.0}    &  {\scriptsize22.2}   \\ 
\hline 
                         \label{tabla1} 
                     \end{tabular} 
      \end{center} 
\end{table*}

\begin{table*} 
      \begin{center} 
            \caption{Journal of mm observations of the \object{GRB 050509b} field.} 
                     \begin{tabular}{@{}lccccc@{}} 
 
{\scriptsize Date of 2005 UT} & {\scriptsize Configuration} & {\scriptsize Detection level (mJy, 1$\sigma$)}  & {\scriptsize Frequency (GHz)} & {\scriptsize Beam (arcsec)}   & {\scriptsize Position angle (deg)}  \\ 
%& {\scriptsize (mJy, 1$\sigma$)} &   {\scriptsize (GHz)}   & {\scriptsize (arcsec)} & {\scriptsize angle (deg)} \\  
%{\scriptsize Date of 2005 UT} & {\scriptsize Configuration} & {\scriptsize Detection level}  & {\scriptsize Frequency} & {\scriptsize Beam}   & {\scriptsize Position}  \\ 
%                &               & {\scriptsize (mJy, 1$\sigma$)} &   {\scriptsize (GHz)}   & {\scriptsize (arcsec)} & {\scriptsize angle (deg)} \\  
\hline 
{\scriptsize May 11, 02:56-03:38} & {\scriptsize 6Dp} &  {\scriptsize 1.9}  &  {\scriptsize 80.327} &  {\scriptsize 32.2 x 4.44} &  {\scriptsize -136} \\
                   &     & {\scriptsize 12.0}  & {\scriptsize 242.842} &  {\scriptsize 9.89 x 1.62} &  {\scriptsize -137} \\
{\scriptsize May 11, 21:38-23:49} & {\scriptsize 6Dp} &  {\scriptsize 0.8}  &  {\scriptsize 80.467} &  {\scriptsize 9.56 x 4.51} &  {\scriptsize + 66} \\
                   &     &  {\scriptsize 5.9}  & {\scriptsize 242.842} &  {\scriptsize 3.24 x 1.73} &  {\scriptsize + 64} \\
{\scriptsize May 13, 00:26-02:44} & {\scriptsize 6Dp} &  {\scriptsize 1.0}  &  {\scriptsize 80.467} & {\scriptsize 25.30 x 4.54} &  {\scriptsize + 48} \\
                   &     &  {\scriptsize 4.4}  & {\scriptsize 221.501} & {\scriptsize 6.05 x 1.70} & {\scriptsize -128} \\
{\scriptsize May 16, 17:33-22:28} & {\scriptsize 5Dp} &  {\scriptsize 0.3}  &  {\scriptsize 92.682} &  {\scriptsize 6.55 x 5.44} &  {\scriptsize + 93} \\
\hline 
                         \label{tabla2} 
                     \end{tabular} 
      \end{center} 
\end{table*}

The {\it Swift} satellite (Gehrels  et al. 2004) offers unique capabilities
for  the detection  of  GRBs thanks  to  its high  sensitivity and  imaging
capabilities at $\gamma$-rays,  X-rays and optical wavelengths. The 
short \object{GRB
050509b}  was discovered by  {\it Swift}/BAT  detector on  9 May  2005. The
burst started at  04:00:19.23 UT and lasted for $\approx40$~ms, 
putting it in the ``short-duration'' class of GRBs.   It had a 
%peak flux of 1.57 $\pm$ 0.36  photons   cm$^{-2}$  s$^{-1}$  and   a  
fluence  of  (0.95  $\pm$ 0.25)$\times10^{-8}$  erg  cm$^{-2}$  in 
the 15-150 keV  range  (Gehrels et al. 2005).  
The prompt dissemination  (13.7 s) of the GRB position (Hurkett
et al. 2005)  enabled prompt responses of automated  and robotic telescopes
on  ground,  like ROTSE-III  (Rykoff  et  al.   2005), RAPTOR  (Wozniak  et
al. 2005) and BOOTES-1 (shown in this paper), 
although no prompt  afterglow emission was detected.  By the time
when {\it Swift} slewed and started  data acquisition (about 60 s after the
event onset), a fading X-ray  emission was detected by the {\it Swift}/XRT;
this can  be considered as the first  clear detection of an afterglow 
in a short duration GRB (Kennea et  al.  2005). 
This triggered a multiwavelength
campaign at  many observatories aimed  at detecting the afterglow  at other
wavelengths, as in the case of the long duration GRB class.
Here we report the results of the multiwavelength observations carried out
by our group, from millimetre (mm) wavelengths to the optical bands.

\section{Observations and data reduction} 
  \label{observaciones} 
  
  The BOOTES-1 very wide field camera, located at Estaci\'on de
  Sondeos Atmosf\'ericos (INTA-CEDEA) in Huelva (Spain), observed the
  region of the sky containing the {\it Swift}/BAT error box of GRB
  050509b as part of its routine observing schedule (Castro-Tirado 
  et al. 2004). A 30~s
  exposure started at 04:00:00 UT (19 s prior to the onset of the
  40 ms short burst), with the following frame starting at 04:01:00 UT.
  A limiting (unfiltered, airmass 4.0) magnitude of 6.0 is derived
  for any prompt optical flash arising from this event.
 
  ToO observations in the optical were triggered starting 0.53~hr after the
  event  at  the  1.2~m  Mercator   telescope  (+~MEROPE  CCD  camera)  at
  Observatorio   del  Roque  de   los  Muchachos   in  La   Palma  (Spain).
  Subsequently,  ToO  observations were  made  at  the  1.5~m telescope  at
  Observatorio de  Sierra Nevada  in Granada (Spain)  and at the  6.0~m BTA
  telescope (+~SCORPIO) at  the Special Astrophysical Observatory (SAO-RAS)
  in  Nizhnij-Arkhyz  (Russia), and at the 2.2~m telescope (+BUSCA) at 
  Calar Alto (Spain). Near infrared  (nIR)  observations  were
  obtained at  the 3.5~m telescope (+~OMEGA2000) at Calar  Alto  as
  part   of   the  ALHAMBRA\footnote{http://alhambra.iaa.es:8080}   back-up
  programme.  Table \ref{tabla1} displays the observing log.

  Additionally, mm observations were  obtained  at  the  
  Plateau de  Bure
  Interferometer (PdBI)  as part of  our ToO programme.  The  PdBI observed
  the source on  May 11 and May 13 (6 antennas compact D configuration) 
  and May 16 (5  antennas compact D configuration). The 
  data reduction was done
  with the standard  CLIC and MAPPING software distributed  by the Grenoble
  GILDAS group\footnote{http://www.iram.es/IRAMFR/GS/gildas/gildas.html}; 
  the flux calibration is relative to the carbon star MWC349.
  Table \ref{tabla2} displays the observing log.
  
  In order to  subtract in all of our optical and  nIR images the 
  contribution
  of  the bright elliptical  galaxy 9$\farcs5$ away from the XRT
  error   box,   we   modeled   it   using  the   ELLIPSE routine under
  IRAF\footnote{IRAF  is distributed  by the  NOAO, which  are  operated by
  USRA,  under  cooperative agreement  with  the  US  NSF.}. We  used  the
  residual  to perform further  analysis in  the optical and nIR
  images.  The optical  field was calibrated using the field 
  photometry provided by Henden (2005). The nIR images were calibrated 
  using  the 2MASS Catalogue.
 
\section{Results and discussion} 
  \label{resultados}

  The main observational result is  the lack of any variable optical/nIR/mm
  counterpart  in  our images, within the refined {\it Swift/XRT} error 
  box (Gehrels et al. 2005), in  spite  of  the intensive  searches,  in
  agreement with the upper limits reported by ROTSE-III and RAPTOR and {\it
  Swift}/UVOT (Gehrels et al. 2005) for the prompt optical emission and 
  by PAIRITEL (Bloom  et al. 2005)  for the prompt nIR  emission.  Similarly,
  the deep upper  limits reported at the Keck (Bloom  et al. 2005; Cenko  
  et al.  2005)  and  at the  VLT  (Hjorth et  al.  2005a; Covino  et
  al.  2005) are  in agreement  with our  conclusions drawn  from  the deep
  6.0BTA observations.  
 
\begin{figure} 
\begin{center}
      \resizebox{6.0cm}{!}{\includegraphics{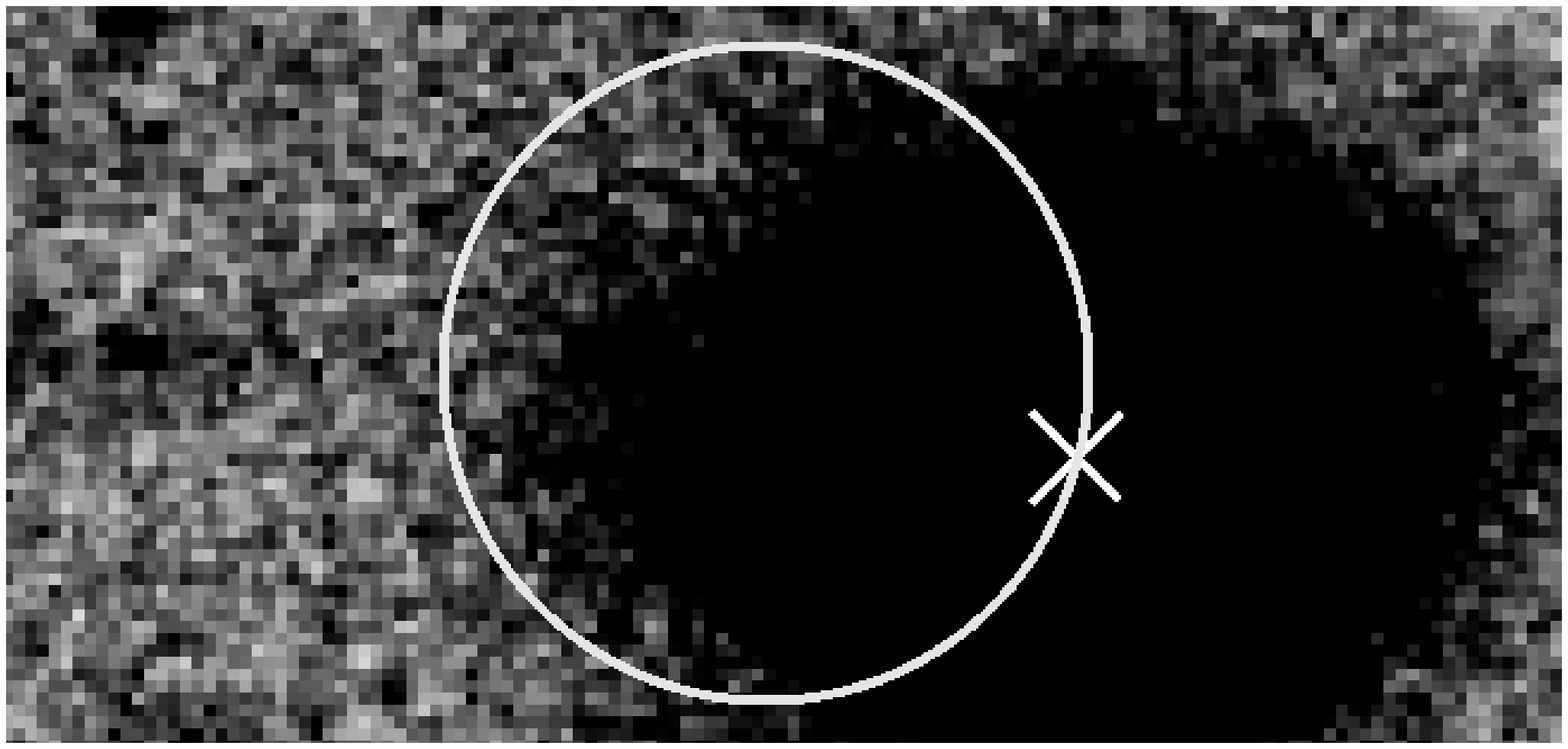}} 
      \resizebox{6.0cm}{!}{\includegraphics{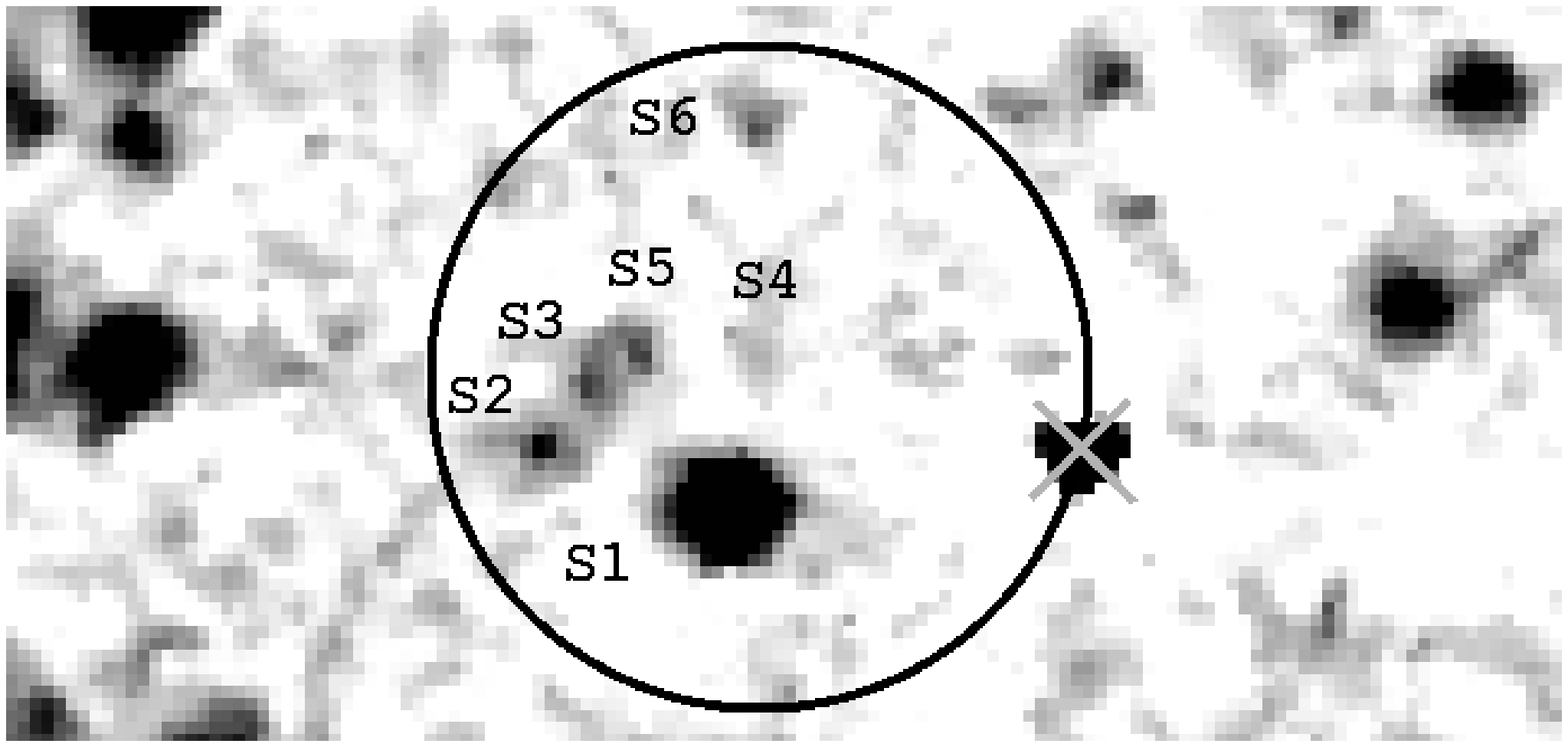}} 
      \caption{The deep  $R$ band image  of the \object{GRB  050509b} field
               taken at the 6.0BTA on  11 May 2005. The contribution of the
               elliptical galaxy,  marked with a cross  ({\it upper panel})
               has been removed in order  to better show the content of the
               {\it  Swift}/XRT error  box  ({\it lower  panel}).  The  six
               sources S1-S4 (Cenko  et al. 2005) and S5-S6 (Hjorth
               et  al.  2005b) within  the  {\it  Swift}/XRT  error box  are
               indicated.    The   field   is   $41^{\prime\prime}   \times
               20^{\prime\prime}$ with North up and East to the left.}
    \label{carta ID}
\end{center} 
\end{figure}

    \subsection{No optical/nIR/mm afterglow at all ?} 
             \label{sin emision} 
              
             The  fact  that  the  detected X-ray  afterglow  for  GRB
             050509b  (Gehrels et al.   2005) is  the faintest  of all
             afterglows  detected by  the {\it  Swift}/XRT so  far may
             indicate that the density of the surrounding medium where
             the progenitor took place  is much lower than the typical
             value  of  $\approx$  1  cm$^{-3}$  derived  for  several
             long-duration GRBs.  Assuming that the X-ray afterglow is
             caused  by synchrotron emission,  as in  the case  of the
             long-duration  family,   one  should  also   expect  some
             contribution at nIR and optical wavelengths, but a simple
             S$_{\rm  X-ray}$/S$_{\rm optical}$ scaling  would predict
             prompt  optical  fluxes $\sim$  10$^{2}$  (i.e. $\sim$  5
             magnitudes) fainter than  the optical afterglows observed
             so far for the long-duration events.
             Moreover, the combined  {\it Swift}/XRT and {\it Chandra}
             X-ray  Observatory  ({\it  CXO}) observations  (Patel  et
             al. 2005)  imply that the decay exponent  $\alpha$ is 1.1
             (Gehrels et  al. 2005), i.e., an  optical afterglow might
             have  rapidly  decayed   in  brightness  with  a  similar
             power-law decay  index for $\nu_{\rm  opt}$ $<$ $\nu_{X}$
             $<$ $\nu_{c}$ (Sari et al. 1998).

        \subsection{Is GRB 050509b at $z$ = 0.225 ?} 

        \label{cluster}                 

        GRB 050509b is  located in the direction of  the NSC J123610+285901
        cluster of  galaxies (Gal  et al.  2003) at $z$  = 0.225  (see also
        Bloom et al.  2005). In fact, during the  prompt search for the
        X-ray afterglow detected by {\it Swift}/XRT, {\it CXO} has detected
        the diffuse  X-ray emission from  the intracluster gas  rather than
        the point source itself (Patel et al. 2005).

        If  this would  be  the  case, taking  into  account the  proximity
        (9$\farcs5$) of the  X-ray  error box  to the  ellip\-tical
        galaxy 2MASX  J12361286+2858580 (Fig. 1), 
        the  relationship to it  cannot be
        discarded. It would be only $\sim$ 33 kpc in  projection (3.563
        kpc/$^{\prime\prime}$), as the angular  size distance $D_{A}$ = 735
        Mpc,  considering a  Hubble constant  of H$_{0}$  = 71  km s$^{-1}$
        Mpc$^{-1}$, a matter density $\Omega_{m}$ = 0.3, and a cosmological
        constant $\Omega_{\Lambda}$  = 0.7.  Assuming  this, the progenitor
        would  have been originated  in the  halo of  the elliptical  galaxy 
        and
        could  favour a neutron-star  merger origin  (Goodman et  al. 1987;
        Eichler et al. 1989; Narayan et al. 1992), a physical scenario that
        can explain a short-duration burst  like GRB 050509b.  We note that
        past  searches for  correlations between  clusters of  galaxies and
        GRBs  did  not  reveal  positive  results  (Hurley  et  al.   1997;
        Gorosabel \& Castro-Tirado 1997).

\begin{figure}
 \begin{center} 
 \resizebox{7.5cm}{5.0cm}{\includegraphics{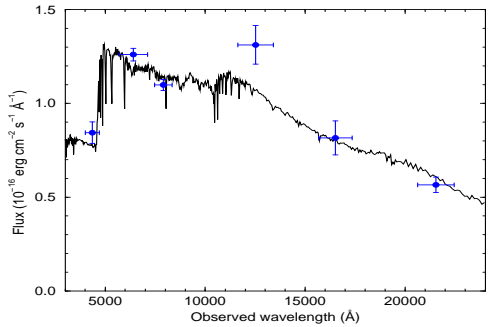}} 
 \caption{Synthetic  fit to  the $BRIJHK$  band photometric  points  of the
 elliptical galaxy. The best fist  is achieved with a moderately 
 extincted ($A_{\rm v}=0.4$  mag) galaxy harbouring an evolved stellar 
 population (age $\sim$360 Myr).  
 The  SED fit  assumed a  Salpeter (1955)  IMF  and a
 Calzetti et al. (2000) extinction law.}
  \label{hostsed}
  \end{center} 
\end{figure}

        Radio emission in (or close to)  the center of this galaxy has been
        detected  at the  WSRT  (van der  Horst  et al.  2005) although  no
        emission lines are  seen in the optical spectrum  (Bloom et al.
        2005).   The  restframe   colours  (and  therefore  the  associated
        K-corrections)  have  been  obtained   based  on  the  HyperZ  code
        (Bolzonella et  al. 2000).  Fitting synthetic spectral energy 
        distributions (SED)  templates to the
        $B$  band   magnitude (19.73 $\pm$ 0.08), derived from the Bok 
        telescope  i\-mage (Engelbracht \& Eisenstein 2005) and  
        from our own data and taking
        the nIR magnitudes  from the 2MASS, produces $U-B$  = $-$0.50 $\pm$
        0.20, $B-V$  = 1.08 $\pm$ 0.20,  $V-R$ = 0.31 $\pm$  0.20 and $R-I$
        =0.69  $\pm$  0.20,  (correcting  all for  the  Galactic  reddening
        E($B-V$) = 0.019; Schlegel et al. 1998).  At $z$ = 0.225, this 
        is a rather  luminous  galaxy, with  M$_{B}$  =  $-$20.6  and 
        M$_{R}$ =  $-$22.2. According to HyperZ, the  $BRIJHK$ band  SED  
        of  the neighbour  galaxy  at $z$ = 0.225 favours 
        ($\chi^{2}$/d.o.f = 2.4) a moderately extincted  
        ($A_{\rm v}\sim  0.4$ mag) galaxy  harbouring an
        evolved     dominant    stellar    population  (age $\sim360$
        Myr). Figure~\ref{hostsed} shows the SED of the elliptical 
        galaxy.

        In this scenario, an intriguing possi\-bility arises 
        if the event would  be the result of  a stellar collapse, similarly  
        to the long duration  GRBs.  At such  a  redshift, an  underlying  
        type Ib/c SN similar to SN 1998bw/GRB 980425 (Galama et al. 1998)  
        should have peaked at $R$ $\sim$ 21, about 20  days since the burst 
        onset. An underlying Type Ia SN  is also expected if the event is the
        result of the gravitational collapse of a C/O white dwarf into a 
        neutron star (Dar  \& De R\'ujula 2004). 
        Our optical limits in the R-band 
        18.5 days after the event onset imply that the peak flux of any 
        underlying supernova should have been $\sim$ 3 magnitudes fainter than
        the one observed for the type Ib/c SN 1998bw/GRB 980425 (Galama et al. 
        1998), and 2.3 magnitudes fainter than a typical type Ia SN 
        (Filippenko 1997 and references therein), in agreement with the VLT 
        results (Hjorth et al. 2005a).

    \subsection{Is GRB 050509b at high redshift ?} 
        \label{high-z}         
        It  is also plausible  that GRB  050509b has  been originated  at a
        redshift considerably higher  than 0.225, i.e. it could  lie in the
        S1 galaxy or in one  of the fainter sources (S2-S6) detected within
        the  XRT error box  (Fig. 1) at unknown redshift (see Bloom et
        al. 2005) or it occurred in a  much more
        distant  object,  at  very  high  redshift.  In  the  former  case,
        extinction might have played a considerable role in order to hide
        optical variability in  the first hours/days following  the event. 
        However,
        the  lack of  detection  of  a nIR  transient  in the  observations
        presented here disfavours this argument. On the contrary,  if GRB
        050509b arose  from a high-redshift host galaxy,  it would 
        have easily  been beyond the limit  of the  
        optical/nIR telescopes,  because  of  the Lyman~$\alpha$  blanketing  
        affecting the  optical band.

\section{Conclusions} 
  \label{conclusiones} 
   
  We  have  shown  multiwavelength   observations  of  the  short  duration
  gamma-ray burst  detected by  {\it Swift} (\object{GRB  050509b}) between
  0 s  and  $\sim$  18.8  days after  the  event.  
  No optical/nIR/mm
  afterglow emission  has been  detected, in spite  of the  reported X--ray
  afterglow  detection by  {\it Swift}  few minutes  after the  event,
  confirming the elusiveness of the afterglow of the short duration events.
  The spectral energy distribution of the neighbouring, potential host galaxy,
  favours a system harbouring an  evolved dominant stellar
  population  (age $\sim360$  Myr), unlike most long duration GRB host 
  galaxies observed so far, i.e. thus giving support to a compact binary 
  merger origin. Any underlying supernova that could be associated 
  with this particular event should have been at least 3 magnitudes fainter 
  than SN 1998bw and 2.3 magnitudes fainter than a type Ia SN.

  GRB 050509b  is the second short  duration GRB which is  detected by {\it
  Swift}/BAT after  GRB 050202 (which occurred too close to the  Sun 
  and could not be properly followed-up), and  the first one localised 
  with high accuracy
  by {\it Swift}/XRT.  In spite of {\it CGRO}/BATSE  detecting about 1/4 of
  events  be\-longing to  the  short  duration class,  {\it  Swift} has  only
  detected  2 (out  of $\sim$ 40 events), most likely due to its softer
  threshold  energy.   However,  thanks  to  its  extraordinary  repointing
  capabilities,  the  accurate  localisations  for future  events  and  the
  corresponding  multiwavelength follow-up,  will  shed more  light on  the
  origin of the short-duration GRBs.
 
\begin{acknowledgements} 
   
  We acknowledge the comments from the anonymous referee, and C.W. 
  Engelbracht and D.J. Eisenstein for making available their B-band image. 
  This  work is  based partly  on observations  ca\-rried out  with  the IRAM
  Plateau de Bure Interferometer, suppor\-ted by INSU/CNRS (France),
  MPG (Germany) and  IGN (Spain); and on data
  products from  the Two Micron  All Sky Survey  (2MASS), which is  a joint
  project of  the Univ.  of  Massachusetts and the IPAC/CalTech,  funded by
  NASA and  the NSF.  The work  was partially supported  by the grant A3003206 
  of the Grant Agency of the Academy of Sciences of the Czech Republic, 
  by the  RFBR grant 04-02-16300a  and  by  the  Program  of the  Presidium  
  of  RAN  entitled
  "Non-stationary Phenomena  in Astronomy  2005'' and by the Spanish 
  MEC programmes  AYA2004-01515 and  ESP2002-04124-C03-01.

\end{acknowledgements}


\begin{thebibliography}{} 
 

\bibitem[]{} Bloom, J.S., Prochaska, J.X., Pooley, D., et al. 2005, ApJ, submitted (astro-ph/0505480)
\bibitem[]{} Bolzonella, M., Miralles, J.-M. \& Pell\'o, R. 2000, A\&A 363, 476  
\bibitem[]{} Calzetti, D., Armus, L., Bohlin, R.C.,  et al. 2000, ApJ  533, 682
\bibitem[]{} Castro-Tirado, A.J., Castro Cer\'on, J.M., Gorosabel, J.,  et al. 2002, A\&A 393, L55
\bibitem[]{} Castro-Tirado, A.J., Jel\'{\i}nek, M., Mateo Sanguino, T.J., et al. 2004, AN 325, 679
\bibitem[]{} Cenko, S.B., Soifer, B.T., Bian, C.,  et al. 2005, GCN Circ. 3401 
\bibitem[]{} Covino, S., Israel, G.L., Antonelli, L.A., et al. 2005, GCN Circ. 3413 
\bibitem[]{} Dar, A. \& De R\'ujula, A. 2004, Physics Reports, 405, 203 
\bibitem[]{} Eichler, D., Livio, M., Piran, T. \& Schramm, D.N. 1989, 
             Nature 340, 126
\bibitem[]{} Engelbracht, C. W. \& Eisenstein, D.J. 2005, GCN Circ. 3420
\bibitem[]{} Filippenko, A.V. 1997, ARA\&A, 35, 309 
\bibitem[]{} Fishman, G.J. \& Meegan, C.A. 1995, ARA\&A, 33, 415 
\bibitem[]{} Gal, R.R., de Carvalho, R.R., Lopes, P.A.A., et al. 2003, AJ 125, 2064
\bibitem[]{} Galama, T.J., Vreeswijk, P.M., van Paradijs, J., et al. 1998, Nature 395, 670
\bibitem[]{} Gehrels, N., Chincarini, G., Giommi, P.,  et al. 2004, ApJ 611, 1005 
\bibitem[]{} Gehrels, N., Barbier, L., Barthelmy, S.D., et al. 2005, Nature, submitted (astro-ph/0505630)
\bibitem[]{} Goodman, J., Dar, A. \& Nussinov, S. 1987, ApJ 314, L7
\bibitem[]{} Gorosabel, J. \& Castro-Tirado, A.J. 1997, ApJ 483, L83
\bibitem[]{} Gorosabel, J., Andersen, M.I., Hjorth, J.,  et al. 2002, A\&A 383, 112
\bibitem[]{} Henden, A. A. 2005, GCN Circ. 3454
\bibitem[]{} Hjorth, J., Sollerman, J., Gorosabel, J.,  et al. 2005a, ApJ, submitted (astro-ph/0506123)
\bibitem[]{} Hjorth, J., Sollerman, J., Jensen, B.L.,  et al. 2005b, GCN Circ. 3410
\bibitem[]{} Hurkett, C., Rol, E., Barbier, L.,  et al. 2005, GCN Circ. 3381 
\bibitem[]{} Hurley, K., Hartmann, D., Kouveliotou, C.,  et al. 1997, ApJ 479, L113
\bibitem[]{} Hurley, K., Berger, E., Castro-Tirado, A.J.,  et al. 2002, ApJ 567, 447
\bibitem[]{} Kehoe, R., Akerlof, C., Balsano, R., et al.  2001, ApJ 554, L159
\bibitem[]{} Kennea, J.A., Burrows, D.N., Housek, J.,  et al. 2005, GCN Circ. 3383 
\bibitem[]{} Klotz, A., B\"oer, M. \& Atteia, J.-L. 2003, A\&A 404, 815
\bibitem[]{} Kouveliotou, C., Meegan, C.A., Fishman, G.J.,  et al. 1993, ApJ, 413, 101
\bibitem[]{} Mazets, E.P., Golenetskii, S.V., Ilyinskii, V.N., et al. 1981, A\&SS, 80, 119
\bibitem[]{} Narayan, R., Paczy\'nski, B. \& Piran, T., 1992, ApJ 395, L83
\bibitem[]{} Patel, S., Kouveliotou, C., Burrows, D.N., et al. 2005, GCN Circ. 3419
\bibitem[]{} Rykoff, E.S., Swan, H., Schaefer, B., et al. 2005, GCN Circ. 3382
\bibitem[]{} Salpeter, E.E., 1955, ApJ 121, 161.
\bibitem[]{} Sari, R., Piran, T. \& Narayan, R. 1998, ApJ 497, L17 
\bibitem[]{} Schelgel, D. J., Finkbeiner, D. P. \& Davis, M. 1998, 
ApJ 500, 525 
\bibitem[]{} van der Horst, A.J., Wiersema, K., Wijers, R.A.M.J., et al. 2005, GCN Circ. 3405
\bibitem[]{} van Paradijs, J., Kouveliotou, C. \& Wijers, R.A.M.J. 2000, ARA\&A, 38, 379.
\bibitem[]{} Wozniak, P., Vestrand, W.T., Wren, J., et al. 2005, GCN Circ. 3414 
 
\end{thebibliography}
\end{document}